\documentclass[12pt]{article}
\usepackage{times}
\usepackage{geometry}
\usepackage[utf8]{inputenc}
\usepackage{enumitem,amssymb}
\usepackage{ragged2e}

\usepackage{graphicx}
\usepackage{hyperref}
\usepackage{color}
\usepackage{natbib}
\usepackage{aas_macros}
\newlist{thematic}{itemize}{8}
\setlist[thematic]{label=$\square$}
\usepackage{pifont}
\newcommand{\cmark}{\ding{51}}%
\newcommand{\done}{\rlap{$\square$}{\raisebox{2pt}{\large\hspace{1pt}\cmark}}%
\hspace{-2.5pt}}

\begin{document}
\raggedright
\huge
Astro2020 Science White Paper \linebreak
Science Platforms for Resolved Stellar Populations in the Next Decade \linebreak
\normalsize

\noindent \textbf{Thematic Areas:} \hspace*{60pt} $\square$ Planetary Systems \hspace*{10pt} $\square$ Star and Planet Formation \hspace*{20pt}\linebreak
$\square$ Formation and Evolution of Compact Objects \hspace*{31pt} $\square$ Cosmology and Fundamental Physics \linebreak
  $\square$  Stars and Stellar Evolution \hspace*{1pt} \done Resolved Stellar Populations and their Environments \hspace*{40pt} \linebreak
  $\square$    Galaxy Evolution   \hspace*{45pt} $\square$             Multi-Messenger Astronomy and Astrophysics \hspace*{65pt} \linebreak
  
\textbf{Principal Author:}

Name: Knut A.G. Olsen	
 \linebreak						
Institution: National Optical Astronomy Observatory 
 \linebreak
Email: kolsen@noao.edu
 \linebreak
Phone: 520-318-8555 
 \linebreak
 
\textbf{Co-authors:} Melissa Graham (U. Washington \& LSST, melissalynngraham@gmail.com), Dara Norman (NOAO, dnorman@noao.edu), St{\'e}phanie Juneau (NOAO, sjuneau@noao.edu), and Adam Bolton (NOAO, abolton@noao.edu)
  \linebreak

\textbf{Abstract  (optional):}
Over the past decade, research in resolved stellar populations has made great strides in exploring the nature of dark matter, in unraveling the star formation, chemical enrichment, and dynamical histories of the Milky Way and nearby galaxies, and in probing fundamental physics from general relativity to the structure of stars.  Large surveys have been particularly important to the biggest of these discoveries.  In the coming decade, current and planned surveys will push these research areas still further through a large variety of discovery spaces, giving us unprecedented views into the low surface brightness Universe, the high surface brightness Universe, the 3D motions of stars, the time domain, and the chemical abundances of stellar populations.  These discovery spaces will be opened
by a diverse range of facilities, including the continuing Gaia mission, imaging machines like LSST and WFIRST, massively multiplexed spectroscopic platforms like DESI, Subaru-PFS, and MSE, and telescopes with high sensitivity and spatial resolution like JWST, the ELTs, and LUVOIR.  We do not know which of these facilities will prove most critical for resolved stellar populations research in the next decade. We can predict,  however, that their chance of success will be maximized by granting use of the data to broad communities, that
many scientific discoveries will draw on a combination of data from them, and that advances in computing will enable increasingly sophisticated analyses of the large and complex datasets that they will produce.  We recommend that Astro2020 1) acknowledge the critical role that data archives will play for stellar populations and other science in the next decade, 2) recognize the opportunity that advances in computing will bring for survey data analysis, and 3) consider investments in Science Platform technology to bring these opportunities to fruition.

\pagebreak
\section{Guiding Questions}
As recognized in several Astro2020 Science White Papers \citep[e.g.,][]{Bechtol19,Simon19,Weisz19,Dey19}, some of the most important questions to be addressed by surveys of resolved stellar populations in the coming decade are:
\begin{itemize}
    \item What is the nature of dark matter? \\
    Resolved stellar populations trace the density and distribution of dark matter over orders of magnitude in halo mass, with faint dwarf galaxies and microlenses being particularly important probes  
    \citep[e.g.,][]{2019arXiv190201055D}.

    \item What are the formation histories of nearby galaxies? \\
    The star formation, chemical enrichment, and dynamical histories of galaxies are encoded in their stars, and in the Milky Way and nearby galaxies may be retrieved by photometric, spectroscopic, and astrometric measurements of resolved stellar populations   
    \citep[e.g.][]{Weisz19}.

    \item How can resolved stellar populations serve as probes of fundamental astrophysics? \\
    Precise measurements of resolved stellar populations probe the physics of stellar interiors \citep{2012A&A...543A..54A}, general relativity and black hole physics \citep[e.g.,][]{1998ApJ...509..678G}, and the progenitor stars of explosions 
    (e.g., \citealt{2011Natur.480..348L,2015PASA...32...16S}).
\end{itemize}
Exploiting the surveys that address these questions will crucially depend on the availability and development of not only data archives but also ``Science Platforms" that include high-level analysis tools, as we will argue below.

\section{Where will the answers come from?}

\subsection{The low surface brightness Universe}
Starting with SDSS, surveys have excelled at illuminating the low surface brightness Universe.  The SDSS Field of Streams \citep{2006ApJ...642L.137B} sparked profound interest in using the halo to understand the accretion history of the Galaxy, while the faint dwarf galaxies found in SDSS \citep[e.g.,][]{2005AJ....129.2692W} both exposed our severely incomplete knowledge of the low-mass galaxy population and engendered the Missing Satellites problem.  The Dark Energy Survey yielded a treasure trove of 17 dwarf satellites \citep{2016MNRAS.460.1270D} and 11 stellar streams \citep{2018ApJ...862..114S}, and introduced the possibility that the Magellanic Clouds represent an infalling group with a dwarf satellite system of its own \citep{2017MNRAS.465.1879S}. 
The GD-1 stellar stream, when isolated by a combination of Pan-STARRS photometry and Gaia DR2 proper motions, shows the clearest evidence yet of a gap from a potential perturbation from a dark matter subhalo \citep{2018ApJ...863L..20P}. {\em All} of these discoveries relied on data archives to selectively filter large catalogs and isolate populations of interest.

In the coming decade, LSST should reveal 100+ dwarf galaxies in the Galactic halo \citep{2018MNRAS.479.2853N} and, by analogy, a similar number of new streams. Spectroscopic surveys with DESI, 4MOST, WEAVE, Subaru-PFS, or MSE will give context of chemistry and dynamics through massive datasets, while ELTs can probe their dark matter profiles through dynamics of their cores.  While any one of these future facilities will have enormous impact on our knowledge of the low surface brightness Universe, the last decade indicates that the greatest discovery and understanding will come from a combination of measurements and datasets and the application of innovative computational techniques to them, areas where data archives will play a critical role.

\subsection{The moving Universe}
The Gaia mission has produced spectacular results on the structure of our Galaxy \citep[e.g.,][]{2018Natur.563...85H}.  While many of these were to be expected from a survey containing parallaxes and proper motions of a billion stars, others came as complete surprises.  Among the totally unexpected results was the discovery of a clear spiral pattern in phase space in the Galactic disk from Gaia DR2 \citep{2018Natur.561..360A}, demonstrating that the disk is not in complete dynamical equilibrium.  Indeed, the pattern is likely the result of the recent passage of a perturbing mass, perhaps that of the Sagittarius dwarf galaxy \citep{2018MNRAS.481.1501B}, raising the possibility of future discoveries of gravitational perturbations in the disk.  Another surprising result was the discovery of a gap in the main sequence luminosity function in the Gaia DR2 100 pc parallax sample \citep{2018ApJ...861L..11J}, which
\citep{2018MNRAS.480.1711M} explain as caused by $^3$He mixing in the convective envelope of stars in a narrow range of mass on the main sequence.  The result demonstrates how precise measurements of large samples can yield new insight into fundamental astrophysics.  The Gaia Archive (https://gea.esac.esa.int/archive/) made all of these discoveries possible through its catalog query and data services.

While Gaia will remain the premier astrometric mission for many years to come, by the end of the next decade LSST will provide parallaxes out to $\sim$300 pc for stars too faint for Gaia, yielding an inventory of all brown dwarfs and white dwarfs in the solar neighborhood; proper motions from LSST will probe the dynamics of the halo out to $\sim$100 kpc with main sequence stars \citep{LSSTSB}.  The LSST Science Platform will be crucial for identifying and analyzing these objects from the 37 billion objects and 7 trillion individual detections in the final catalog (https://www.lsst.org/scientists/keynumbers).

\subsection{The high surface brightness Universe}
The components of galaxies that contain the bulk of the stellar mass -- nuclei, bulges, disks, and main bodies --  are typically high in surface brightness and thus severely crowded.  Surveys with HST have played a crucial role in deciphering the star formation and chemical enrichment histories of galaxies in the Local Volume \citep[e.g. ANGST;][]{2009ApJS..183...67D}, including that of M31 \citep[PHAT;][]{2012ApJS..200...18D}.  The availability of these datasets through the MAST Archive have made them broadly used and highly cited.  In the Galactic center, adaptive optics-corrected instruments on large ground-based telescopes have permitted the observation of complete orbits of stars around the Milky Way's central black hole \citep[e.g.,][]{2017PhRvL.118u1101H}, and the detection of a flare from a star near the black hole, whose time delays are a direct probe of general relativity \citep{2018A&A...618L..10G}.  Because the timescales of the stellar orbits span decades, data archives have proved crucial in supporting these discoveries.

In the next decade, surveys with e.g. JWST and the ELTs will make deconstruction of the star formation histories of galaxies across the Hubble sequence possible \citep[e.g.,][]{2003AJ....126..452O}.  The high resolution of AO-corrected ELTs will permit the detection of faint stars close to the central black hole, bringing tests of general relativity into a new regime \citep{2017arXiv171106389D}.  Data archives should support these observations by making the data broadly accessible, exposing the analysis software to the public, and linking the observations over time.

\subsection{Time-resolved Surveys}
Surveys targeting time-variable phenomena are providing powerful perspectives on resolved stellar populations.  For example, \citet{2019MNRAS.482.4562B} used RR Lyrae from Gaia DR2, 2MASS, CRTS, and Pan-STARRS
to associate the Virgo Stellar Stream with the Magellanic Stream.  Extragalactic time-domain studies in resolved stellar populations inform us about the late-stages of stellar evolution and SN progenitor stars \citep[e.g., luminous blue variables;][]{2015MNRAS.447..598S}, and supply temporally- and spatially-resolved SN light echoes which provide a unique and valuable view of historical Galactic explosions (e.g., \citealt{2008ApJ...681L..81R}, and see also Astro2020 paper \citealt{MLGWP}).  All of these studies relied on data archive catalog and image services.

The LSST's final catalogs will contain $\sim$135 million variable stars \citep{LSSTSB}, a much larger sample of fainter objects with longer duration light curves (up to 10 years) in multiple filters ({\it ugrizy}) compared to past surveys. LSST will recover 50\% of the RR Lyrae stars at 600 kpc, adding great detail to our map of the Milky Way, its halo streams, and dwarf satellites \citep{2012AJ....144....9O}.
Beyond the Milky Way, 
Cepheid variables will provide distances to more SN\,Ia host galaxies, improving both cosmological analyses \citep[e.g.,][]{2016ApJ...830...10H} and our physical understanding of the explosions \citep[e.g.,][]{2018arXiv180608359F}. 
The future wide-field time-domain datasets will be large and continuously evolving. Ingesting and processing 10 million LSST alerts per night will require dedicated community resources in order to extract features from light curves and/or cross-match to archival catalogs for historical context, and promptly prioritize objects for follow-up (e.g., spectroscopy).

\subsection{Spectroscopic Surveys}
The line-of-sight velocities and chemical abundances from stellar spectra provide crucial insights to the study of the star formation, chemical enrichment, and dynamical histories of the Milky Way and nearby galaxies.  The concept that the unique chemical signatures of clusters of stars should permit us to unscramble the early formation history of the Galaxy \citep{2002ARA&A..40..487F} provided motivation for the sample of $\sim$1 million stars collected by several surveys (SDSS I-IV, APOGEE, LAMOST, GALAH, RAVE).  These surveys have demonstrated the ability of chemical tagging to pick out known objects purely in abundance space \cite[e.g.,][]{2016ApJ...833..262H,2018MNRAS.473.4612K}, and have led to the development of significant new abundance measurement techniques \citep[e.g. The Cannon; The Payne;][]{2015ApJ...808...16N,2018arXiv180401530T} and theoretical frameworks for interpreting the surveys \citep[e.g.,][]{2017Galax...5...43S}. 

The spectroscopic surveys of the next decade (SDSS-V, 4MOST, DESI, WEAVE, and Subaru-PFS) will yield samples at least 10$-$100$\times$ larger than the current ones. In combination with the Gaia mission, these samples will construct a multi-dimensional map of the Milky Way \citep{Dey19}, allowing for detailed exploration of the dark matter distribution, Galactic substructure, and rare objects like the most metal-poor stars.  Mining these complex datasets will be greatly served by the ability of data archives and Science Platforms to access, analyze, visualize, and model the data.

\section{Recommendations for Archives and Science Platforms}
Surveys from the facilities and missions planned or proposed for the coming decade will open tremendous opportunities for using resolved stellar populations to answer fundamental astrophysical questions.  
It is critical that the data from these future surveys be made broadly accessible and reusable.  The 20-year SDSS survey has resulted in 7700 refereed papers containing SDSS data, the large majority from authors not part of the SDSS Collaboration (https://www.sdss.org/science/). Papers based on archival use of HST data outpace those based on GO programs \textcolor{black}{(376 {\it vs.} 341 in 2017; \citealt{HSTPUBSTAT})}.  In one year, Gaia DR2 has resulted in $\sim$300 refereed publications, $\sim$500 if arXiv preprints are included.

The tremendous productivity of archival data use clearly shows that data archives have a critical role to play for resolved stellar populations and all other science in the next decade, as archives make data broadly accessible and lower the barrier for entry into astronomy research.  {\bf We recommend that Astro2020 acknowledge this critical role of data archives.}  It also indicates that there is potential for still greater exploitation of archival data.  Several institutions are developing Science Platforms to expand the range of services for large survey datasets, including running complex workflows on servers close to the where the data reside; some currently operating examples of Science Platforms for astronomy are SDSS SkyServer (http://skyserver.sdss.org), SciServer (http://www.sciserver.org), NOAO Data Lab (https://datalab.noao.edu), and MAST Labs (https://mast-labs.stsci.io), soon to be joined by the LSST Science Platform (https://ldm-542.lsst.io).  


Based on the input of participants at the workshop ``NOAO Community Needs for Science in the 2020s", in the next decade Science Platforms should:
\begin{enumerate}
    \item {\bf Serve the Pixels} with built-in tools for pixel-level alignment, analysis, and modeling of images from a variety of multi-wavelength heterogeneous imaging datasets, including pixel-level data access for spectra. 
    \item {\bf Co-locate the Catalogs} of astrometry, photometry, \textcolor{black}{spectroscopic, and time-domain data}, provide built-in tools for fast catalog coordinate cross-matching, and provide functionality to make cross-matching codes and data products widely shareable.
    \item {\bf Capture the Moment} by providing access to public Alerts brokers for capturing time variable phenomena, target observation manager (TOM) infrastructure for following up high value events, and servers for the associated processing and storage. ANTARES (https://antares.noao.edu) is an example of a currently operating public Alert broker, while AEON (http://ast.noao.edu/data/aeon) is TOM infrastructure under development.
    \item {\bf Curate and Preserve} existing data sets indefinitely, along with a platform that contains embedded query, analysis, and visualization tools tailored to the data.
    \item {\bf Model the Data} with algorithms for global parameter fits across all archived data, via e.g.\ platform-embedded Deep Learning, including mixed-origin sources and heterogeneous surveys (imaging, catalogs, and spectra).
\end{enumerate}
{\bf We recommend that Astro2020 recognize Science Platforms as the natural evolution of today’s archives for the next decade of large surveys.} They will let users bring code to the data for those cases where transferring the data is impractical. They will leverage new technological developments from the fast-paced software industry to increase the scientific return of large survey datasets.  Perhaps most importantly, they will make the ability to explore and analyze large datasets available to a broad community, such that the potential for discovery will be limited by imagination rather than technical hurdles.
{\bf We recommend that Astro2020 consider investments in Science Platform development to allow them to evolve the way astronomical research is performed in the next decade.}
Science platforms could be the “Google Office Suite” for astronomical data analysis.  Collectively, they would have all of the major survey catalogs, images that cover most of the sky, millions of spectra, and a full-blown computing environment, all at a user's fingertips. Users could upload their own data and their own software and share whatever they like with collaborators, and see what others do with their data as well. \textcolor{black}{The availability of archived data, combined with standardized formats and co-located analysis tools, may also improve the scientific integrity of our work by making astronomical research more easily reproducible.} Such platforms could unleash innovation in resolved stellar populations and other research as significant as that of new facilites.

\begin{figure}[t]
    \centering
    \vspace{-0.45in}
    \includegraphics[width=1.0\textwidth]{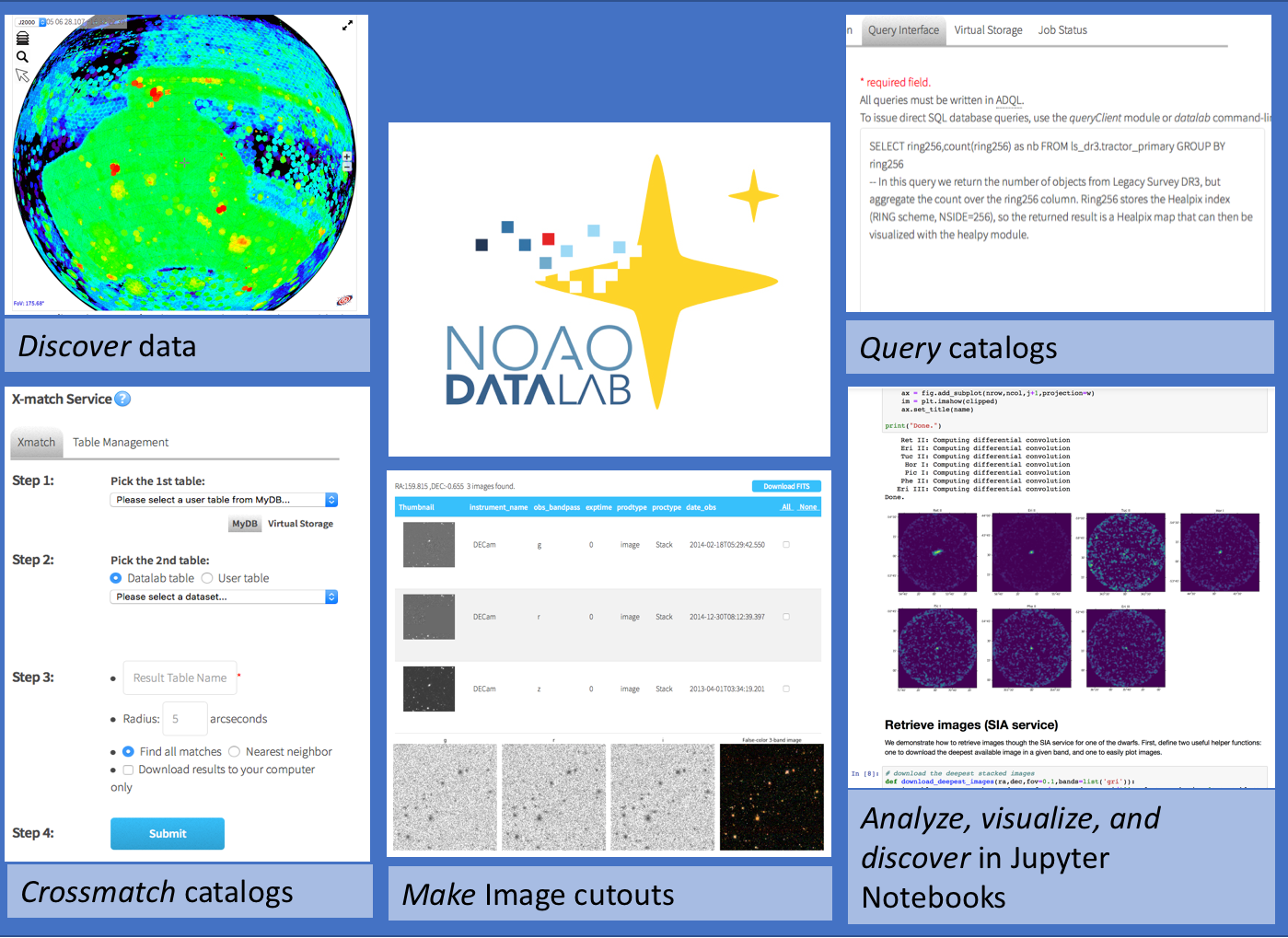}
    \vspace{-0.2in}
    \caption{
    An overview of services provided by the NOAO Data Lab (https://datalab.noao.edu), a prototype Science Platform.  The figure at bottom right is from a Jupyter Notebook that demonstrates the automated detection of low surface brightness dwarf galaxies in the Dark Energy Survey.  Science Platforms aim to enable data discovery, visualization, and analysis close to significant data volumes, provide a convenient way to share data and workflows with collaborators and the community, and make data and analysis software broadly available, useful, and reproducible.
     }
    \label{fig:imaging}
\end{figure}

\pagebreak
\let\oldbibliography\thebibliography
\renewcommand{\thebibliography}[1]{\oldbibliography{#1}
\setlength{\itemsep}{0pt}} 
\bibliography{main}

\end{document}